\documentclass[aps,prd,showpacs,showkeys,superscriptaddress,twocolumn]{revtex4-1}
\usepackage[colorlinks,linkcolor=blue,anchorcolor=blue,citecolor=blue,urlcolor=blue,breaklinks=true]{hyperref}
\usepackage{graphicx}
\usepackage{amsmath}
\usepackage{amssymb}
\usepackage{slashed}
\usepackage{latexsym}
\usepackage{epsfig}
\usepackage{amsbsy}
\usepackage{array}
\usepackage{changes}
\usepackage{amssymb}
\usepackage{setspace}
\usepackage{bm}
\usepackage{lipsum}
\usepackage{mathrsfs}
\usepackage{float}
\usepackage{color}
\usepackage{graphicx}
\usepackage{subfigure}
\usepackage[T1]{fontenc}
\usepackage{mathptmx}
\DeclareMathAlphabet{\mathcal}{OMS}{cmsy}{m}{n}
\DeclareSymbolFont{largesymbols}{OMX}{cmex}{m}{n}

\begin{document}
\author{Xu Wang}
\affiliation{School of Astronomy and Space Science, Nanjing University, Nanjing 210023, People's Republic of China}

\author{Abdusattar Kurban}
\affiliation{Xinjiang Astronomical Observatory, Chinese Academy of Sciences, Urumqi 830011, Xinjiang, People's Republic of China}

\author{Jin-Jun Geng}
\affiliation{Purple Mountain Observatory, Chinese Academy of Sciences, Nanjing 210023, People's Republic of China}

\author{Fan Xu}
\affiliation{School of Astronomy and Space Science, Nanjing University, Nanjing 210023, People's Republic of China}

\author{Xiao-Li Zhang}
\affiliation{Department of Physics, Nanjing University, Nanjing 210093, People's Republic of China}

\author{Bing-Jun Zuo}
\affiliation{Department of Physics, Nanjing University, Nanjing 210093, People's Republic of China}

\author{Wen-Li Yuan}
\affiliation{Department of Physics, Nanjing University, Nanjing 210093, People's Republic of China}

\author{Yong-Feng Huang}
\email{hyf@nju.edu.cn}
\affiliation{School of Astronomy and Space Science, Nanjing University, Nanjing 210023, People's Republic of China}
\affiliation{Key Laboratory of Modern Astronomy and Astrophysics (Nanjing University),
	Ministry of Education, Nanjing 210023, People's Republic of China}
\accepted{by PRD} 

\title{Tidal Deformability of Strange Quark Planets and Strange Dwarfs}

\begin{abstract}
{Strange quark matter, which is composed of u, d, and s quarks, could be the true ground state of
matter. According to this hypothesis, compact stars may actually be strange quark stars, and
there may even be stable strange quark dwarfs and strange quark planets.
The detection of the binary neutron star merger event GW170817 provides us new clues on
the equation of state of compact stars. In this study, the tidal deformability of strange
quark planets and strange quark dwarfs are calculated. It is found to be smaller than that of normal matter counterparts. For a typical
0.6 M$_\odot$ compact star, the tidal deformability of a strange dwarf is about 1.4 times less
than that of a normal white dwarf. The difference is even more significant between strange quark
planets and normal matter planets. Additionally, if the strange quark planet is a bare one
(i.e., not covered by a normal matter curst), the tidal deformability will be extremely small,
which means bare strange quark planets will hardly be distorted by tidal forces. Our study
clearly proves the effectiveness of identifying strange quark objects via searching for
strange quark planets through gravitational wave observations.}
\bigskip

%\pacs{12.38. Lg, 12.38. Mh, 64.60. an}

\end{abstract}

\maketitle

%\Pages{35}{37}

\maketitle

\section{Introduction}
The study of strong interaction at suprasaturation density is an important subject
in modern physics. Ground-based experiments can place stringent constraints on the
equation of state near the nuclear saturation density \cite{2014NuPhA.922....1D,2010ApJ...722...33S}.
However, the density at the center of neutron stars can be even higher than 5 times of the
nuclear saturation density, of which the conditions cannot be imitated in
laboratory \cite{1979ARA&A..17..415B,2004Sci...304..536L}.
We thus have to resort to astronomical observations to probe the internal structure of
compact stars. But due to the inaccuracy of the measurements of mass and radius for them,
the equation of state (EoS) of compact stars is still highly
uncertain \cite{2004Sci...304..536L}.

In 2015,the Laser Interferometer Gravitational-wave Observatory (LIGO) detected the first gravitational wave event GW150914, which
was produced by the merger of two black holes \cite{2016PhRvX...6d1015A,2016PhRvL.116v1101A}.
The masses of the two black holes are 36$_{-4}^{+5}$ M$_\odot$ and 29$_{-4}^{+4}$ M$_\odot$,
respectively \cite{ 2016PhRvL.116x1102A}. Two years later, in 2017, the first gravitational wave
signal from a binary neutron star merger event, GW170817, was detected by the LIGO and Virgo detectors \cite{2017PhRvL.119p1101A}. The total mass of the neutron star binary is  2.74$_{-0.01}^{+0.04}$ M$_\odot$.
A short gamma-ray burst (GRB 170817A) was observed 1.7 s after the coalescence time, which is believed to be
the electromagnetic counterpart of GW170817 \cite{ 2017ApJ...848L..12A,2017ApJ...848L..13A}.
For the gravitational wave emission from binary neutron star mergers, it is interesting that
the tidal effect will change the wave form to some extent, thus it is possible to diagnose
the EoS of neutron stars with the so called tidal deformability \cite{2019JPhG...46l3002G}.
In this way, the observations of GW170817 have provided us with the first constraint on neutron star tidal
deformability, which is an extremely useful feature for studying neutron star
EoS \cite{2018PhRvL.121p1101A,2019PhRvX...9a1001A,2017ApJ...850L..19M, 2018PhRvL.120q2703A,2018PhRvL.120z1103M, 2018PhRvL.121i1102D,2017ApJ...850L..34B,2018ApJ...852L..29R}.

The tidal deformability also could be used to distinguish between strange quark stars and
neutron stars.  According to a longly existed hypothesis, strange quark matter may be the
true ground state of baryon matter, so the observed pulsars may actually be strange quark
stars \cite{1984PhRvD..30.2379F,1984PhRvD..30..272W, 1987PhLB..192...71O,2021PhRvC.103c5806H}. Due to the
special nature of quark stars, they have very different mass - radius relations \cite{1989PhRvL..63.2629G}.
But for a 1.4 M$_\odot$ strange star, its radius is very similar to that of a typical 1.4 M$_\odot$
neutron star, which makes it very difficult to distinguish quark stars from neutron stars
observationally. It is interesting to note that quark stars rely on strong interactions to
restrain themselves rather than solely on gravity. It means that there could be strange quark
planets in the universe \cite{1984PhRvD..30.2379F,1984PhRvD..30..272W,1986ApJ...310..261A}.
Furthermore, it even allows a strange dwarf to exist, in the form of a strange quark core
covered by a normal matter crust \cite{1995ApJ...450..253G,1995PhRvL..74.3519G}.
At the center of a strange quark dwarf, the density will be 100 --- 1000 times larger
than that of normal white dwarfs. Interestingly, some possible phenomena associated with the crust
collapse of strange quark stars have been
studied \cite{1998Sci...280..407C,2004ChA&A..28..144J,2018ApJ...858...88Z,2021Innov...200152G}.

The tidal deformability of strange quark stars has been calculated by several groups \cite{2010PhRvD..82b4016P,2020PhRvD.102b8501T,2020PhRvD.101f3023L}.
For strange stars and neutron stars, this parameter again is
quite similar at 1.4 M$_\odot$. It shows some difference mainly at the low mass end.
We thus need to pay attention to some more special merger events. It has been argued that
the gravitational wave signal from the merger of a strange planet with a strange star in the Milky Way
could be detected by the LIGO detector, and it could even be detected by the next-generation Einstein
Telescope when the merger happens in local galaxies up to several Mpc \cite{2015ApJ...804...21G,2019AIPC.2127b0027K}.
Strange planets and strange dwarfs are hopeful candidates that could be used to test the
strange quark matter hypothesis \cite{2017ApJ...848..115H,2020ApJ...890...41K,2020arXiv201205748K},
but their tidal deformability has not been thoroughly investigated.
Here we will calculate the tidal deformability of strange quark planets and strange dwarfs. Our
study is aimed at merger events of strange star - strange planet systems and strange star - strange
dwarf systems.

This paper is structured as follows. In Section \ref{sec2}, we introduce the tidal deformability and
the method to calculate them. In Section \ref{sec3}, we present our numerical results for strange dwarfs and strange planets. The combined tidal deformability of strange quark binaries
is also calculated. Finally, Section \ref{sec4} is our conclusion and discussion.

\section{Method}\label{sec2}

\subsection{Tidal deformability}

In a strong tidal field, the structure of a compact object will be distorted.
Tidal deformability is used  to measure this distortion, which is defined as the ratio of one star's
induced quadrapole moment $Q_{ij}$ with respect to its companion's tidal field
$E_{ij}$, i.e., \cite{2008PhRvD..77b1502F,1998PhRvD..58l4031T}
\begin{equation}
\lambda=-\frac{Q_{ij}}{E_{ij}}.
\end{equation}
Sometimes we use a dimensionless tidal deformation parameter defined as,
\begin{equation}
\Lambda=\frac{\lambda c^{10}}{G^4m^5},
\end{equation}
where $m$ is the mass of the star, $G$ is the gravitational constant and
$c$ is the speed of light.

The quantities that can be directly measured from gravitational wave observations are
the combined tidal deformability and the chirp mass $\mathcal{M}$, which are related to the tidal
deformability and the mass of each star in the compact binary
as \cite{2017PhRvL.119p1101A,2019PhRvX...9a1001A},
\begin{equation}
\mathcal{M}=\frac{(m_1m_2)^{3/5}}{M^{1/5}},
\end{equation}
\begin{equation}
\widetilde{\lambda}=\frac{1}{26}\left[\frac{m_1+12m_2}{m_1}\lambda_1+\frac{m_2+12m_1}{m_2}\lambda_2\right] .
\label{l1}
\end{equation}
Here, $m_1$ and $m_2$ are the mass of each object, and $M=m_1+m_2$ is the total mass of the
compact star binary. We assume $m_1$ $\textgreater$ $m_2$ and use $q=m_2/m_1$ $\textless$ 1 to
characterize their mass ratio. In Eq.$\eqref{l1}$, $\lambda_1$ and $\lambda_2$ are the tidal deformability of
each object. Similarly, we can define a dimensionless combined tidal deformability
as \cite{2017PhRvL.119p1101A,2019PhRvX...9a1001A}
\begin{equation}
\widetilde{\Lambda}=\frac{16}{13}\frac{(m_1+12m_2)m_1^4\Lambda_1+(m_2+12m_1)m_2^4\Lambda_2}{(m_1+m_2)^5}.
\label{cl}
\end{equation}

A star with a large radius and a small mass will be easily distorted in the tidal field,
so its tidal deformability is large. On the contrary, a compact object with a smaller radius
and a larger mass will be less distorted in the tidal field, so its tidal deformability is
small \cite{2019JPhG...46l3002G}. The tidal deformability is uniquely determined by the EoS.
Combining the EoS and the Tolman-Oppenheimer-Volkoff (TOV) equation, the tidal deformability
can be numerically calculated. On the other hand, the tidal deformability can be inferred
from the gravitational wave signal in the final stage of the merging process.
At this stage, because the distance between the two objects is very small, the distortion
due to the tidal effect from their companion is large, which will finally
affect the phase evolution of the gravitational waveform. Therefore, the EoS of the
object can be hinted by analyzing the gravitational wave signals. The dimensionless tidal
deformability of the 1.4 M$_\odot$ neutron star is constrained to be less than 800 from
the observation of GW170817 \cite{2017PhRvL.119p1101A}. The result has ruled out some
stiff compact star models \cite{2020GReGr..52..109C}.

\subsection{Calculation of the tidal deformability}
The tidal Love number $k_2$ is also a key parameter that describes
the change in the quadrapole of a star in the tidal field. It is related to the
tidal deformability by \cite{1909RSPSA..82...73L,2010PhRvD..81l3016H}
\begin{equation}
\lambda=\frac{2R^5}{3G}k{_2}
\label{td}
\end{equation}
$k_2$ itself can be calculated from
\begin{equation}
 \begin{split}
k_2=\frac{8C^5}{5}(1-2C)^2[2+2C(y-1)-y][2C[6-3y+3C(5y-8)]\\
+4C^3[13-11y+C(3y-2)+2C^2(1+y)]\\
+3(1-2C)^2[2-y+2C(y-1)]\ln(1-2C)]^{-1}.
 \end{split}
 \label{k}
\end{equation}
Here, $R$ is the radius of the star and $C=Gm/Rc^2$ is its compactness parameter.
The quantity $y=y(R)$ is the solution of the following differential equation from the quadrapole
metric perturbation \cite{2010PhRvD..82b4016P,2020PhRvD.102b8501T,2010PhRvD..81l3016H,2021Symm...13..183K},
\begin{equation}
ry\prime(r)+y(r)^2+F(r)y(r)+Q(r)=0,
\label{y}
\end{equation}
where $F(r)$ and $Q(r)$ are give by \cite{2010PhRvD..82b4016P,2020PhRvD.102b8501T,2010PhRvD..81l3016H}
\begin{equation}
F(r)=\frac{1-4\pi Gr^2(\varepsilon(r)-p(r))/c^4}{1-2Gm(r)/(rc^2)}
\end{equation}
and
\begin{equation}
 \begin{split}
Q(r)=\frac{4\pi Gr^2/c^4}{1-2Gm(r)/(rc^2)}\left[5\varepsilon(r)+9p(r)+\frac{\varepsilon(r)+p(r)}{c_s(r)^2}c^2-\frac{6c^4}{4\pi r^2G}\right]\\
-4\left[\frac{Gm(r)/(rc^2)+4\pi r^2Gp(r)/c^4}{1-2Gm(r)/(rc^2)}\right]^2.
 \end{split}
\end{equation}

In the above equations, $p(r)$ and $\varepsilon(r)=\rho(r)c^2$ are pressure and energy density at radius
$r$. $c_s=c\sqrt{dp/d\varepsilon}$ is the sound speed. These equations need to
be solved together with the Tolman-Oppenheimer-Volkoff equation,
\begin{equation}
\frac{\mathrm{d}p(r)}{\mathrm{d}r}=-\frac{G\varepsilon(r)m(r)}{c^2r^2}\left[1+\frac{p(r)}{\varepsilon(r)}\right]
\left[1+\frac{4\pi p(r)r^3}{c^2m(r)}\right]\left[1-\frac{2Gm(r)}{c^2r}\right]^{-1},
 \label{tov1}
\end{equation}
and
\begin{equation}
\frac{\mathrm{d}m(r)}{\mathrm{d}r}=4\pi r^2\rho(r).
 \label{tov2}
\end{equation}
The boundary conditions of this set of equations are $m(0)=0$, $p(0) = p_c$,
$\varepsilon(0) = \varepsilon_c$ and $y(0)=2$ \cite{2010PhRvD..82b4016P,2020PhRvD.102b8501T,2009PhRvD..80h4035D},
where $p_c$ and $\varepsilon_c$ are the pressure and energy density at the center of the star.

Postnikov et al. pointed out that Eq.$\eqref{k}$ will be invalid when $C$ is very small,
like in the case of  strange dwarfs and strange planets \cite{2010PhRvD..82b4016P,2020PhRvD.102b8501T}.
So, when $C$ $\textless$ 0.1, we use a Taylor series of Eq.$\eqref{k}$ to calculate $k_2$ \cite{2010PhRvD..82b4016P,2020PhRvD.102b8501T}, which is
\begin{equation}
\begin{split}
k_2=\frac{(1-2C)^2}{2}[\frac{2-y}{3+y}+
\frac{(-6+6y+y^2)C}{(3+y)^2}\\
+\frac{(12-8y+34y^2+y^3)C^2}{7(3+y)^3}\\
+\frac{(36+48y+84y^2+62y^3+y^4)C^3}{7(3+y)^4}\\
+\frac{5(648+1476y+1884y^2+1472y^3+490y^4+5y^5)C^4}{147(3+y)^5}\\
+o(C^5)].
\end{split}
\end{equation}
When $C$ $\rightarrow$ 0, Eq.$\eqref{y}$ and Eq.$\eqref{k}$ return to the Newtonian case \cite{2008ApJ...677.1216H}, i.e.,
\begin{equation}
ry\prime (r)+y(r)^2+y(r)-6+4\pi Gr^2\frac{\rho}{c_s(r)^2}=0 ,
\end{equation}
and
\begin{equation}
k_2=\frac{1}{2}(\frac{2-y}{3+y}).
\end{equation}

In the case of strange quark objects covered by a nuclear matter crust,
phase transition occurs and a discontinuity of density will exist inside
the star. It will affect the calculation of $k_ 2$.
In this case, $y$ around the discontinuity is \cite{2010PhRvD..82b4016P,2020PhRvD.102b8501T,2009PhRvD..80h4035D}
\begin{equation}
y(r_d^+)-y(r_d^-)=-\frac{\Delta\rho c^2}{m_0c^2/4\pi r_d^3+p_d},
\label{gap}
\end{equation}
where $r_d$ is the position of the discontinuity, $r_d^-$ is the position at surface of
strange quark core and $r_d^+$ corresponds to the bottom of the crust ($r_d^+$ - $r_d$ $\rightarrow$ +0, $r_d$ - $r_d^-$ $\rightarrow$ +0).
In Eq.$\eqref{gap}$, $m_0$ is the mass of the inner strange quark core, $\Delta\rho=\rho(r_d^-)-\rho(r_d^+)$, and $p_{d}$ is the pressure at $r_d$.

For a given equation of state, we can use the above method to solve the TOV function (Eq.$\eqref{tov1}$, Eq.$\eqref{tov2}$)
and the metric function (Eq.$\eqref{y}$) together to get the solution of y(R). Than we substitute $y(R)$ into Eq.$\eqref{k}$
and Eq.$\eqref{td}$ so as to get the tidal love number and the tidal deformability.

\section{Results}\label{sec3}

We use the phenomenological bag model to describe the strange quark matter,
for which the chemical equilibrium is maintained by weak interactions. Assuming the mass
of strange quarks to be zero, the EoS of the bag model is \cite{1984PhRvD..30.2379F}
\begin{equation}
p=\frac{1}{3}(\varepsilon-4B),
\label{bag}
\end{equation}
where $B$ is the bag constant which represents the pressure just below the surface. According to strange quark hypothesis, the bag constant is limited by $145 \rm MeV<B^{1/4}<162 \rm MeV$ ($57 \rm MeV/fm^3<B<90 \rm MeV/fm^3$)\cite{2010LNP...811.....S}. In our calculations,
we assume $B = 57,70,80,90 \rm MeV/fm^3$ to illustrate the effect of the bag constant on
the tidal deformability of strange quark stars.

A strange quark object is consisted of a strange core and a normal matter crust. Since the density of the
crust is lower than the nuclear saturation density, we use the Baym-Pethick-Sutherland (BPS) \cite{1971ApJ...170..299B}
equation of state to describe it. The pressure ($p_d$) at the boundary between the core and the crust is an
important parameter. It determines the values of $\rho(r^-_d)$ and $\rho(r^+_d)$.
$\rho(r^+_d)$ must be smaller than the neutron drip density of 4.3$\times$10$^{11}$ g/cm$^3$,
which surely is the ultimate maximum density that the crust could
reach \cite{1992ApJ...400..647G,1995PhRvL..74.3519G,1995ApJ...450..253G}. In 1997, Huang et al.
pointed out that when the $\rho(r^+_d)$  reaches $\sim 8.3 \times$10$^{10}$ g/cm$^3$,
the gap between the crust and the quark core will be too small to support the
crust \cite{1997A&A...325..189H}. So, in our calculations, we will take  $\rho(r^+_d) = 8.3 \times 10 ^{10}$ g/cm$^3$.

\subsection{Tidal deformability of bare strange quark objects}

\begin{figure}
\centering
\subfigure{\includegraphics[width=7.5cm]{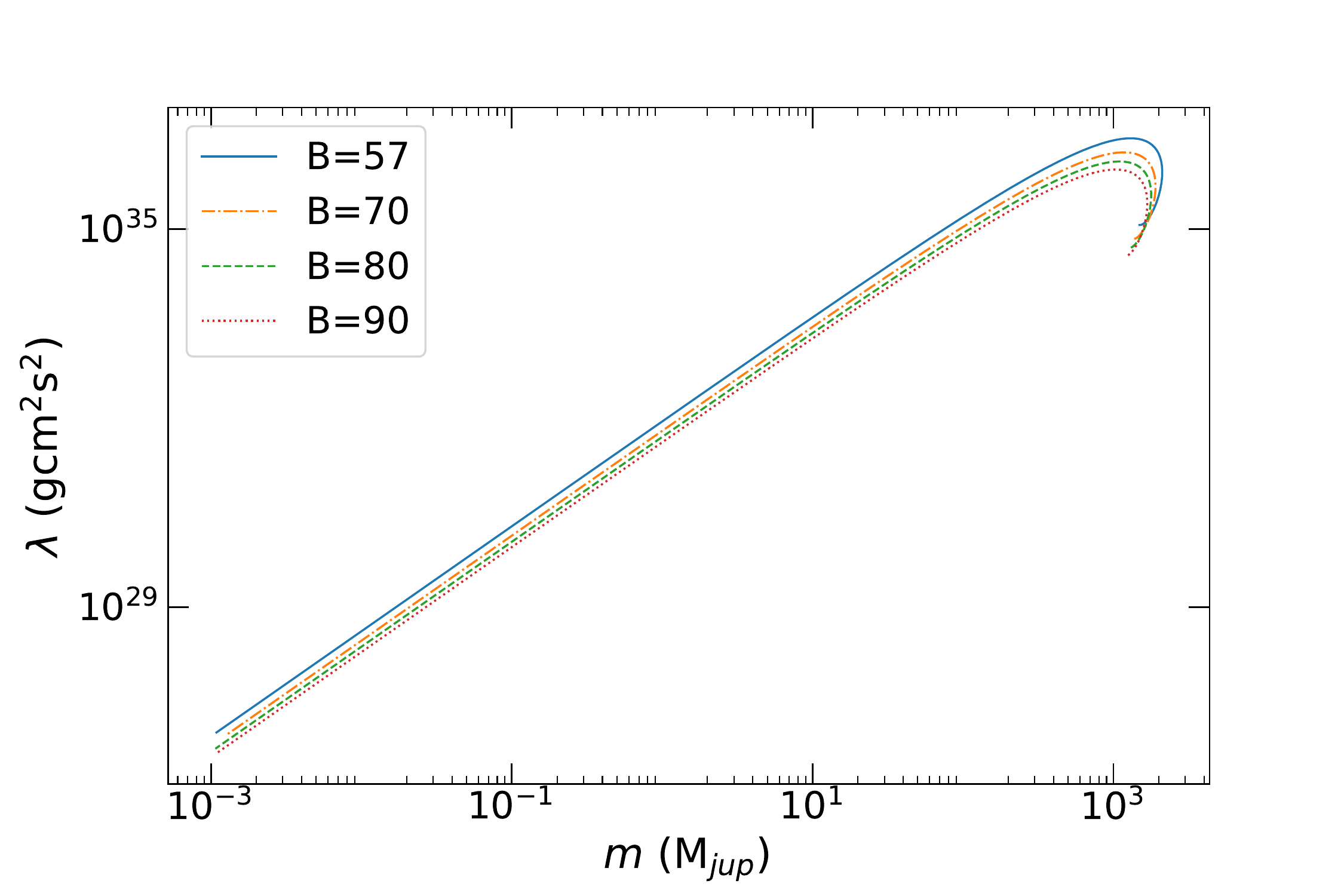}}
\subfigure{\includegraphics[width=7.5cm]{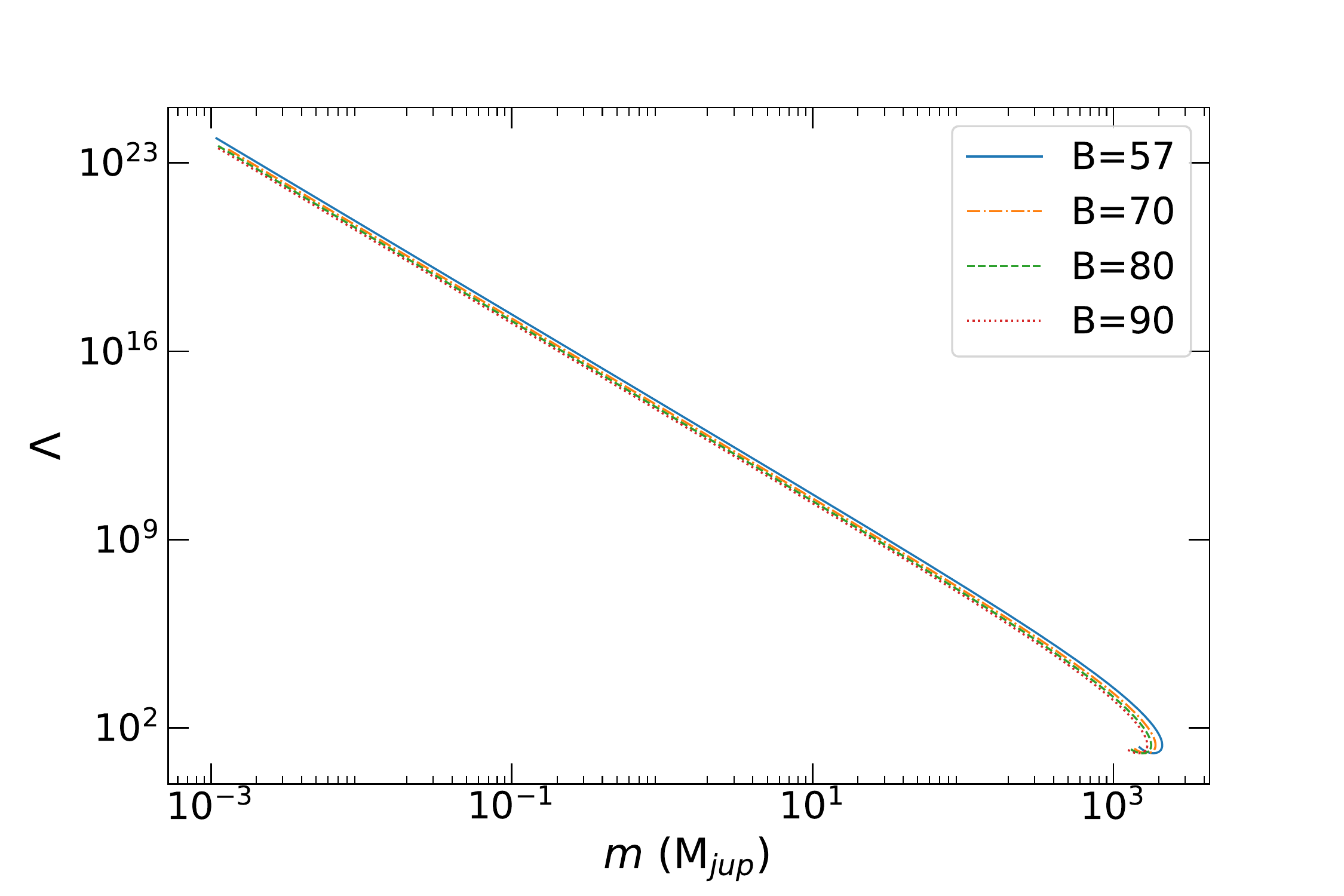}}
\caption{Tidal deformability (the upper panel) and dimensionless tidal deformability (the lower panel)
   versus the stellar mass for bare strange quark stars. Different line styles represent different bag
   constant, which are marked in the figure in units of $ \rm MeV/fm^3$. The stellar mass is in units
   of Jupiter mass (M$\rm_{jup}$). }
\label{fig:s}
\end{figure}

The tidal deformability of strange stars whose masses exceed 0.1 M$_\odot$ has been extensively
studied by several groups \cite{2010PhRvD..82b4016P,2020PhRvD.102b8501T,2020PhRvD.101f3023L}.
In this study, we will mainly focus on the tidal deformability of less massive strange quark
stars, i.e. strange quark planets and dwarfs. Here, we first concentrate on bare strange objects.
Using the EoS described above, we have calculated the structure and the tidal deformability
of bare strange quark stars. In Fig.~\ref{fig:s}, we show the tidal deformability and the dimensionless
tidal deformability as a function of mass. From the upper panel, we see that as the mass increases,
the tidal deformability also increases in most of the mass range. The tidal deformability of a strange
star of 1.4 M$_\odot$ is about 1000 times larger than that of a strange quark planet of about one Jupiter
mass. It means that in the tidal field, bare strange quark planets deform much less than massive bare
strange quark stars. Additionally, when a larger bag constant is assumed, the tidal deformability becomes
smaller. This is also easy to understand, since a larger bag constant generally leads to a larger mean
density, which will make the object less deformed in the tidal force field.
The lower panel of Fig.~\ref{fig:s} illustrates the dimensionless tidal deformability versus the
stellar mass. It is interesting to note that the dimensionless tidal deformability continue to decrease
as the mass increases in most of the mass range. This tendency is very different from that shown in the upper panel.

Inside bare strange quark planets and dwarfs, the density does not vary too much so that it can
be regarded as a constant ($\bar{\rho}$). Then the relationship between the total mass and radius is
\begin{equation}
m=\frac{4}{3}\pi R^3\bar{\rho}.
\label{r1}
\end{equation}
At the same time, the numerical calculation indicates that $k_2$ is also always close to 0.75. Then, the tidal deformability  can be analytically derived as,
\begin{equation}
\lambda=\frac{1}{2G}\left(\frac{3}{4\pi\bar{\rho}}\right)^{\frac{5}{3}}m^{\frac{5}{3}},
\label{r2}
\end{equation}
and
\begin{equation}
\Lambda=\frac{c^{10}}{2G^5}\left(\frac{3}{4\pi\bar{\rho}}\right)^{\frac{5}{3}}m^{\frac{-10}{3}}.
\end{equation}
It means that the theoretical slopes of the curves in the upper and lower panels should be
$5/3$ and $-10/3$, correspondingly. Our numerical results are consistent with these theoretical
expectations.

\subsection{Tidal deformability of strange quark objects with crust}

\begin{figure}
\centering
\subfigure{\includegraphics[width=7.5cm]{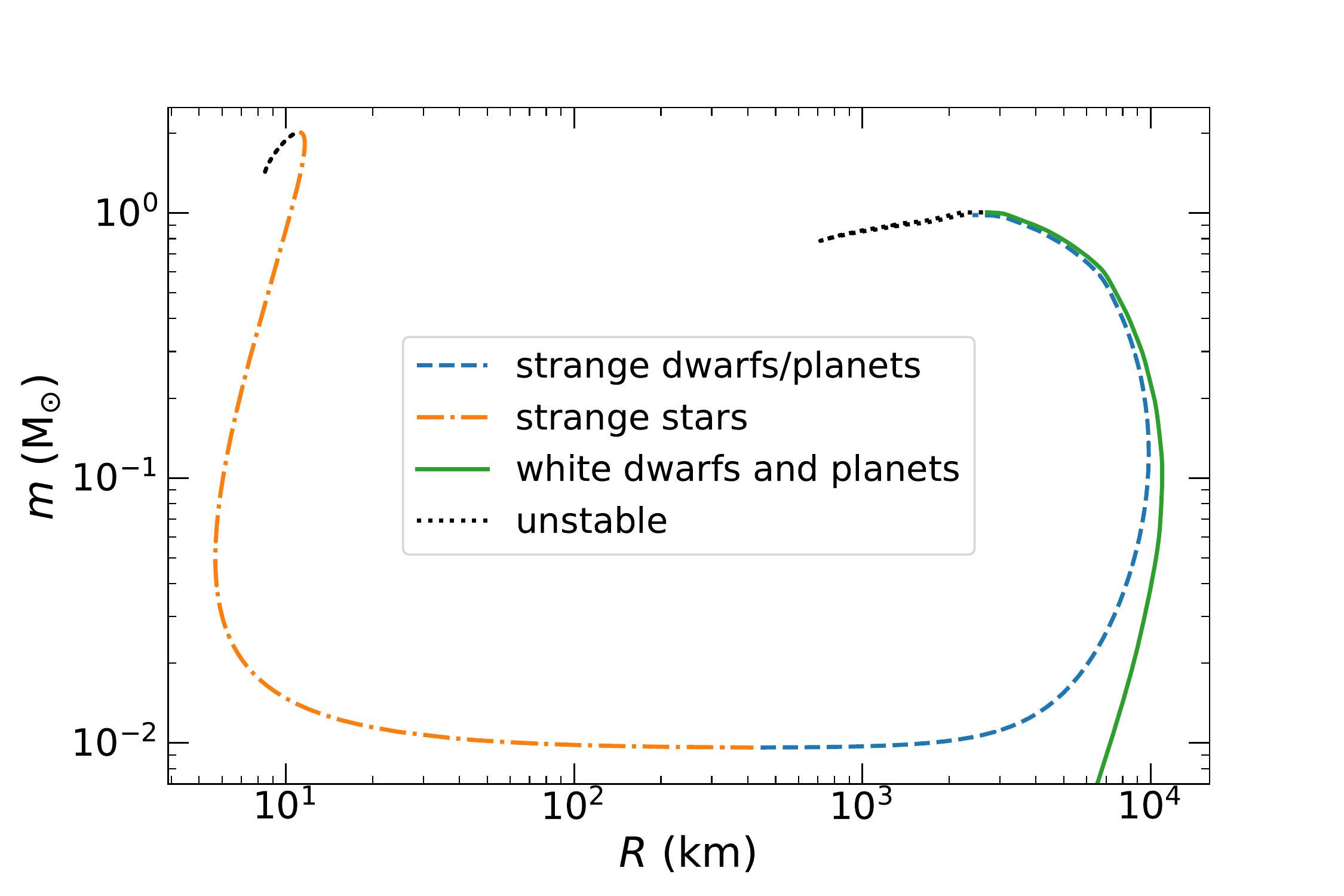}}
\subfigure{\includegraphics[width=7.5cm]{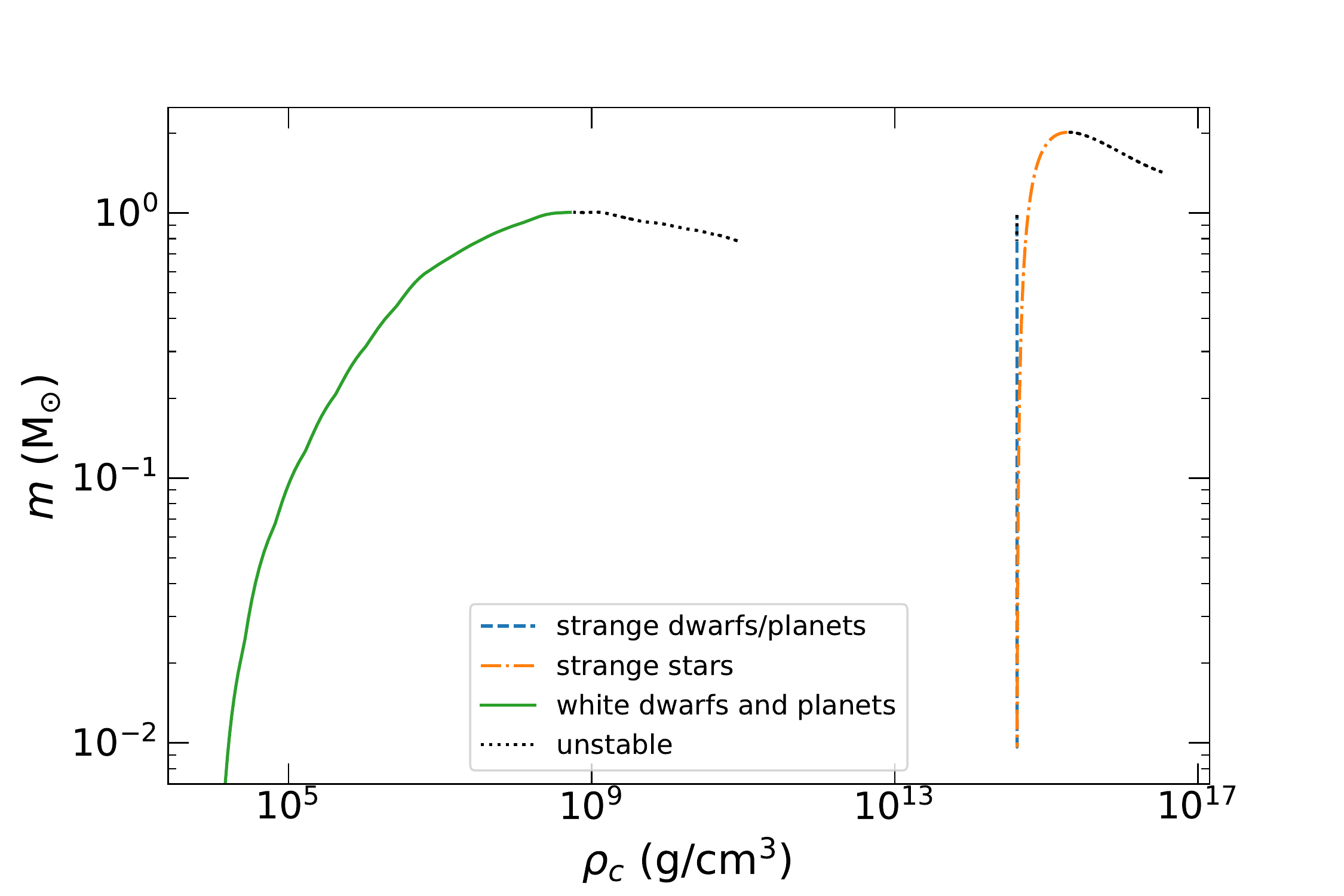}}
    \caption{Stellar mass versus radius (the upper panel) and stellar mass versus
    mass-central density (the lower panel) for the whole sequence of strange quark objects with
    crust. The dash-dotted segment represents strange quark stars (which are the counterparts
    of classical neutron stars). The dashed segment represents strange dwarfs and strange
    planets. The solid segment represents normal white dwarfs and normal planets. The dotted
    segment represents unstable branch. The bag constant is taken as $B$=57 MeV/fm$^3$.}
    \label{fig:sdmr}
\end{figure}

Fig. \ref{fig:sdmr} plots the mass-radius relation and mass-central
density ($\rho_c$) of strange quark stars with crust. The whole sequence includes
strange stars, strange dwarfs and strange planets, compared with normal white dwarfs
and normal planets. Note that those objects in the dotted segment are unstable\cite{1966ApJ...145..505B}.
All other segments of the strange dwarf/strange planet curve are stable against
radial oscillations because of the existence of the quark 
core \cite{1995PhRvL..74.3519G,1995ApJ...450..253G}. In the mass-central density
panel, the left branch corresponds to normal planets and normal white dwarfs, while 
the right branch corresponds to strange quark objects of different masses. We see that 
the central density of strange objects is always larger than $4.1 \times 10^{14}$ g/cm$^3$. 
This is easy to understand. For these objects, whenever there is a strange quark matter 
core, the central density will definitely be higher than $4B$ (see the EoS of Eq. $\eqref{bag}$), which is just the above density value. 

\begin{figure}
\centering
\subfigure{\includegraphics[width=7.5cm]{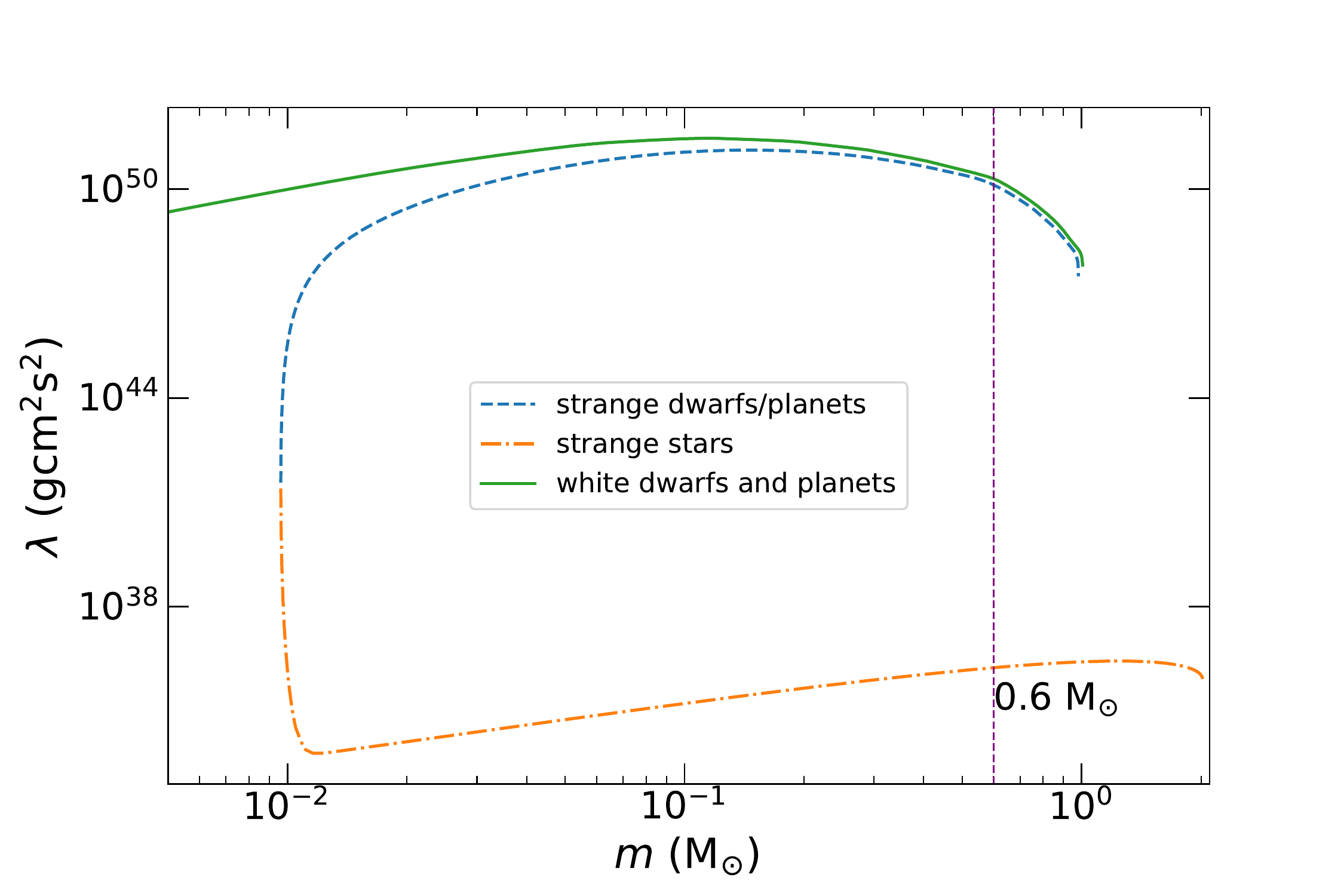}}
\subfigure{\includegraphics[width=7.5cm]{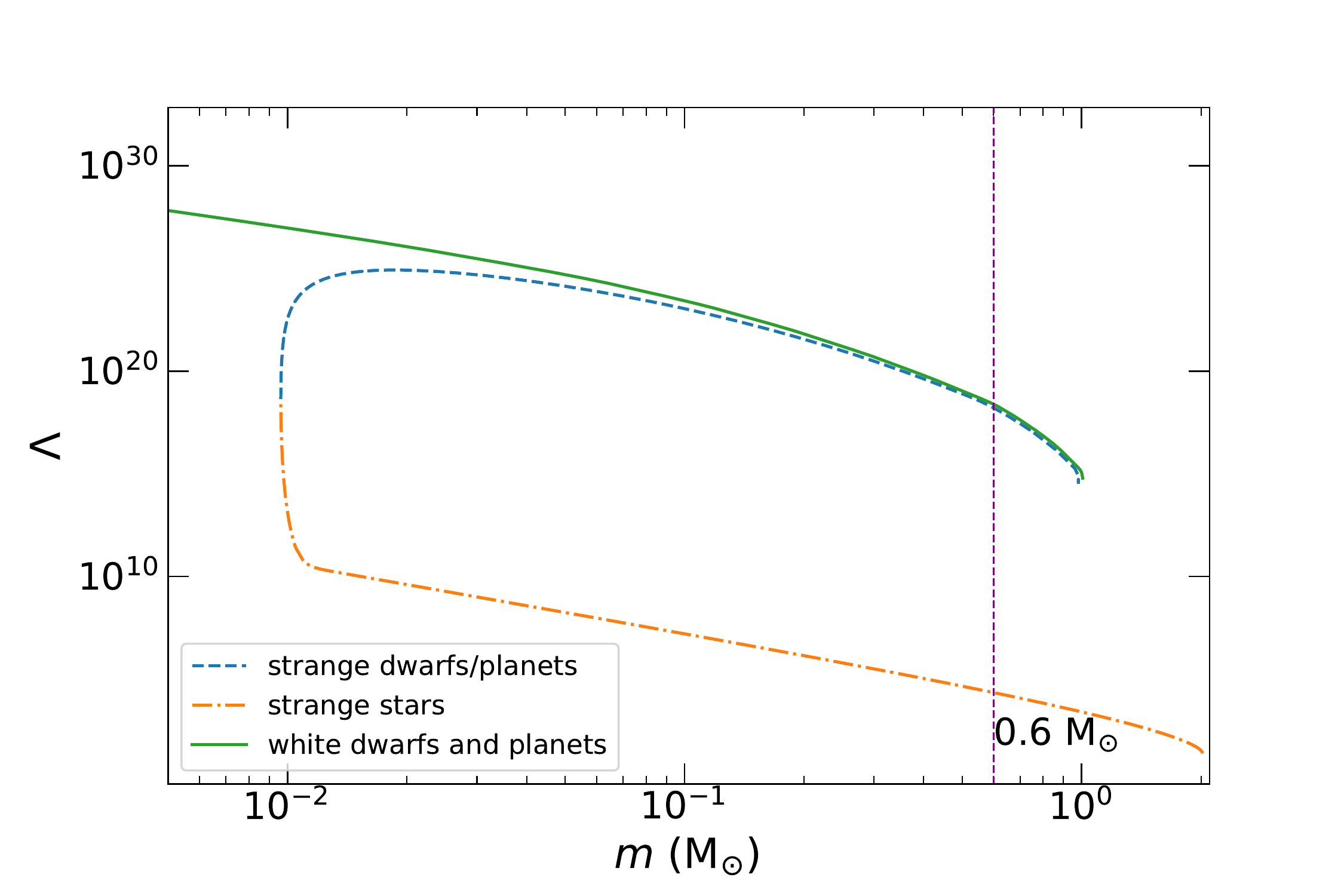}}
\caption{Tidal deformability (the upper panel) and dimentionless tidal deformability (the lower panel)
    of strange quark objects with crust. Line styles and parameters are the same as those in Fig. 2.
    For clarity, the unstable branch is not shown in this plot.}
\label{fig:C}
\end{figure}

We have also calculated the tidal deformability of the whole strange quark object sequence.
The results are shown in Fig.~\ref{fig:C}. Comparing with a 1.0 --- 2.0 M$_{\odot}$ strange
star, a strange dwarf/planet has a smaller mass and a much larger radius. Consequently, it has
a much smaller mean density and will be easily deformed in a tidal field, which means a
strange dwarf/planet will have a much larger tidal deformability as compared with
a 1.0 --- 2.0 M$_{\odot}$ strange star. This can be clearly seen in both panels of
Fig.~\ref{fig:C}. For a strange dwarf, when the mass is larger than $\sim 0.1$ M$_\odot$,
the crust is generally not too thick and it does not significantly affect the tidal
deformability. As a result, the tidal deformability does not differ too much for
a strange dwarf and a normal white dwarf when their masses are larger than 0.1 M$_\odot$.
This is consistent with previous researches by Postnikov et al. \cite{2010PhRvD..82b4016P}. Anyway, we
note that the tidal deformability of strange dwarfs is still generally less than that
of white dwarfs of the same mass. For example, at the typical mass of $\sim$ 0.6 M$_\odot$,
the tidal deformability of a strange dwarf is about 1.4 times less than that of a normal
white dwarf. In short, a strange quark object generally has a smaller tidal deformability
as compared with its normal matter counterpart. Gravitational wave observations thus
can effectively help us identify strange quark objects, especially strange dwarfs/planets \cite{2015ApJ...804...21G,2019AIPC.2127b0027K}.

\subsection{Binary parameters}
The tidal deformability affects the phase evolution of gravitational wave signal. This effect
can be measured by the combined tidal deformability $\widetilde{\lambda}$  \citep{2010PhRvD..81l3016H}.
On one hand, the combined tidal deformability can be directly obtained from gravitational wave
observations \cite{2010PhRvD..81l3016H}. On the other hand, from Eq.$\eqref{cl}$, we see that
the combined tidal deformability can be calculated by considering the tidal deformability and mass
of each object in the compact star binary \cite{2017PhRvL.119p1101A,2019PhRvX...9a1001A}.

In our study, we have calculated the combined tidal deformability of strange star binaries.
For the mass of the primary compact star in the binary system, we have taken three typical
values, i.e., 1.0 M$_\odot$, 1.4 M$_\odot$ and 2.0 M$_\odot$. The companion is assumed to
have a smaller mass, characterized by the mass ratio $q$. When the companion's mass is
very low, it is a strange planet and the binary become a strange star - strange planet
system. We consider the two cases that the strange quark objects are either bare strange
stars or covered with a crust.

\begin{figure}
\centering
\subfigure{\includegraphics[width=7.5cm]{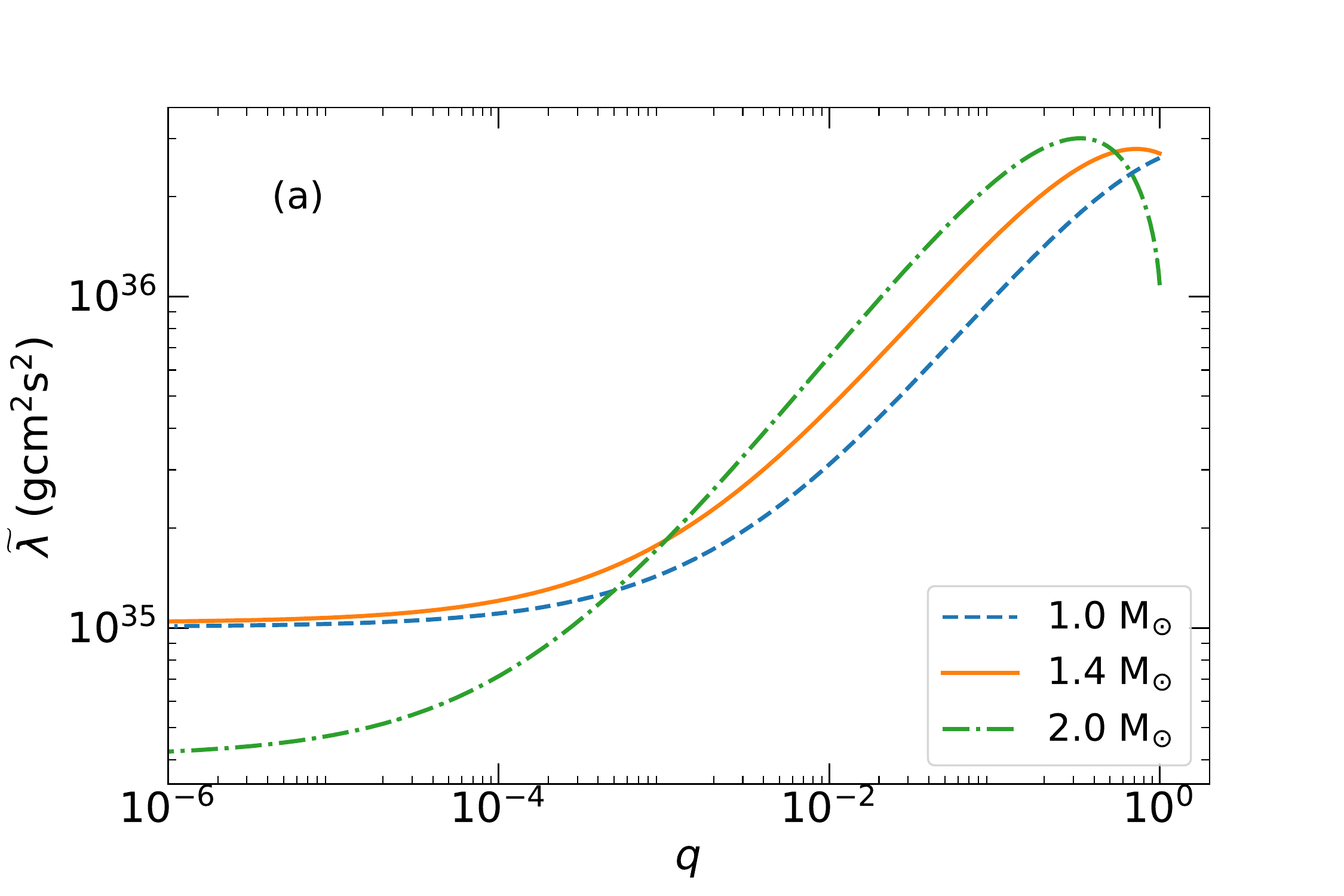}}
\subfigure{\includegraphics[width=7.5cm]{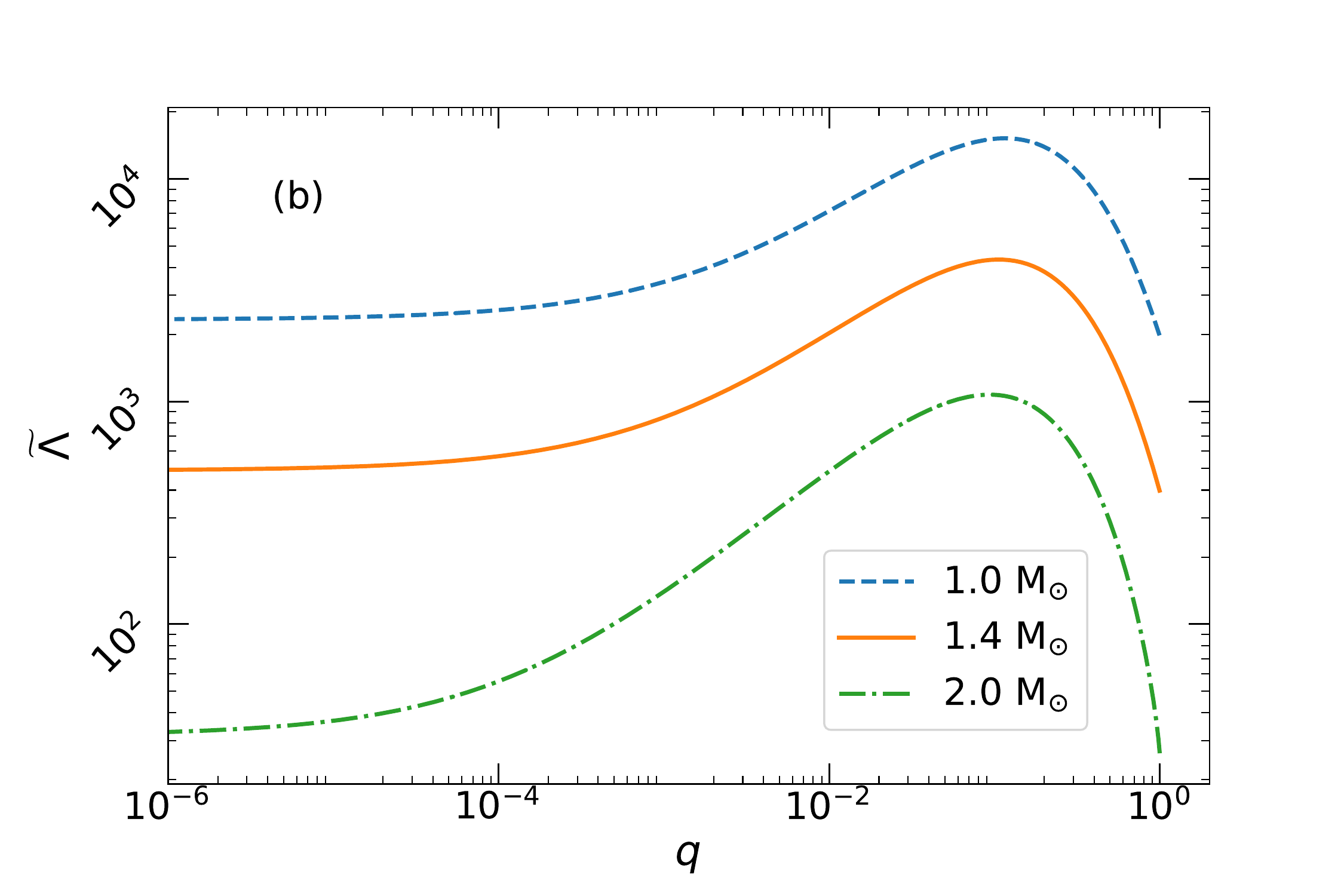}}
\\
\centering
\subfigure{\includegraphics[width=7.5cm]{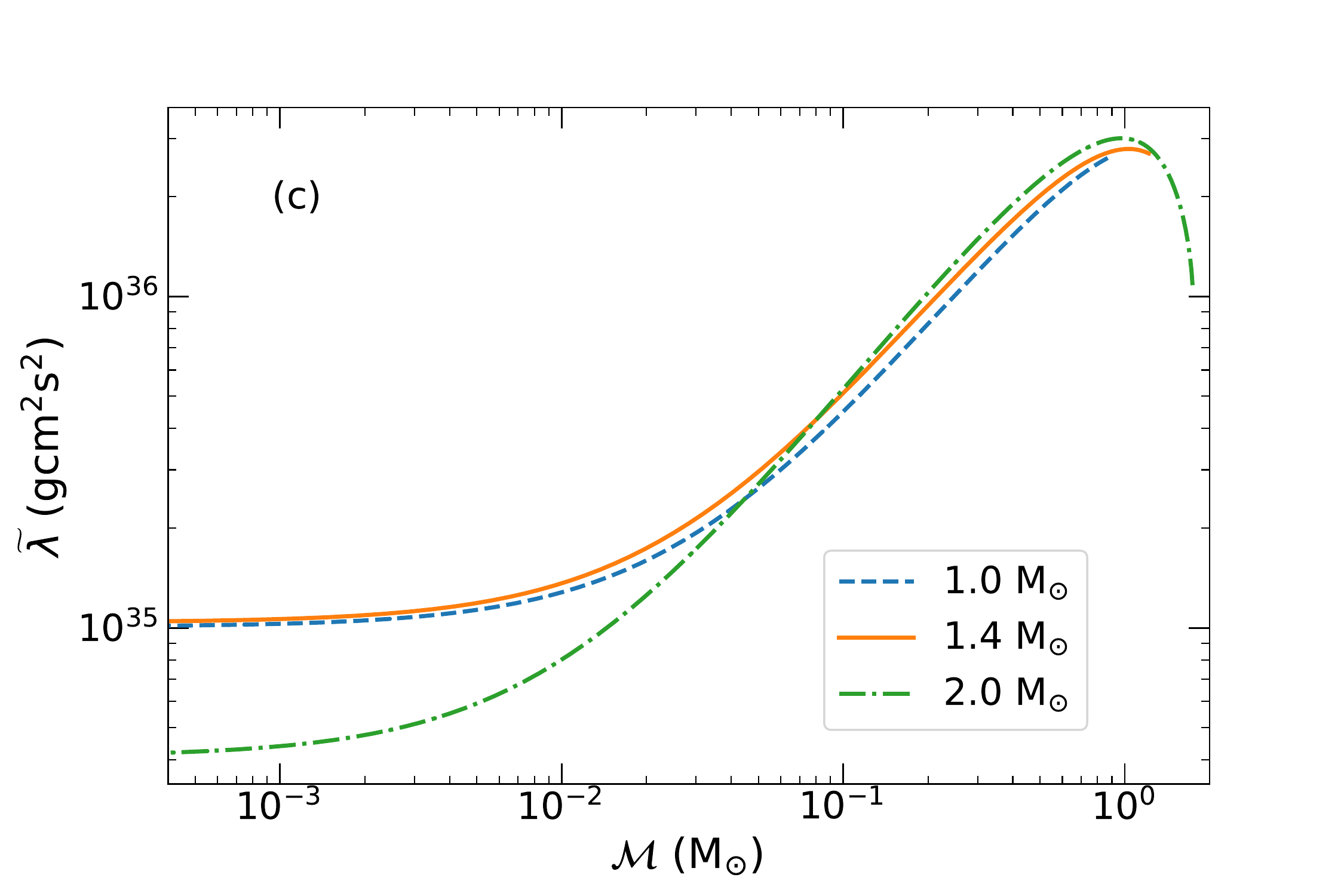}}
\subfigure{\includegraphics[width=7.5cm]{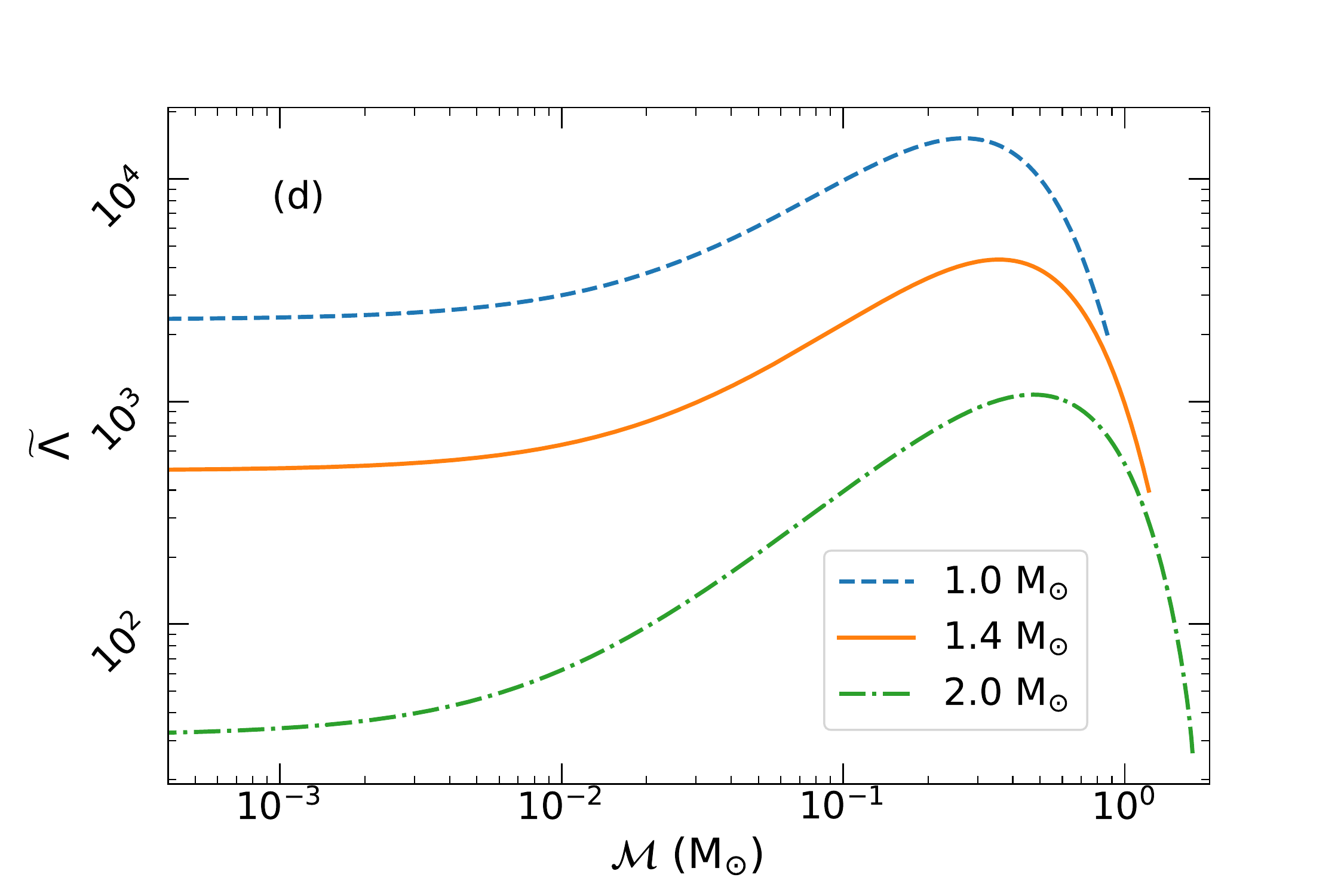}}
\caption{The combined tidal deformability (a) and combined dimensionless tidal deformability (b) as a
   function of the mass radio $q$ for bare strange quark objects. In Panels (c) and (d), the combined
   tidal deformability and combined dimensionless tidal deformability are plot versus
   the chirp mass $\mathcal{M}$. The mass of the primary strange quark star is marked in each panel.}
\label{fig:bs}
\end{figure}

For bare strange quark stars, our numerical results on $\widetilde{\lambda}$ and
$\widetilde{\Lambda}$ are shown in Fig. \ref{fig:bs}.
We see that when the companion becomes less massive, the combined tidal deformability
generally decreases continuously. For the strange star-strange planet systems, the combined tidal
deformability is about $\rm\sim10^{35}\,g\,cm^2\,s^2$, which is much smaller than
that of binary neutron star systems (which is $ \rm\sim 10^{36}\,g\,cm^2\,s^2$). Note that the curves become flat at the low mass-ratio
regime, which means the combined tidal deformability is largely independent of the
planet mass.

\begin{figure}
\centering
\subfigure{\includegraphics[width=7.5cm]{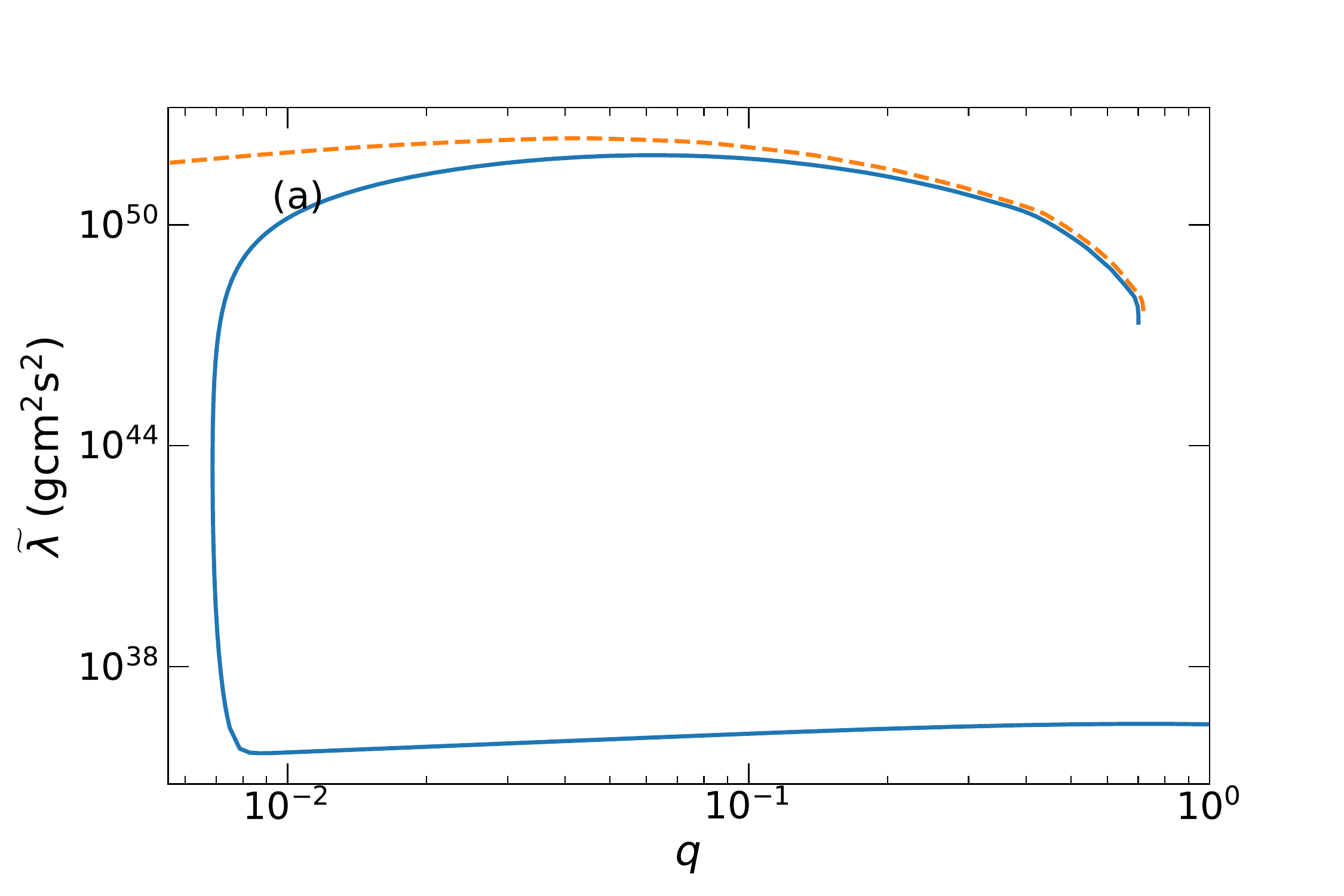}}
\subfigure{\includegraphics[width=7.5cm]{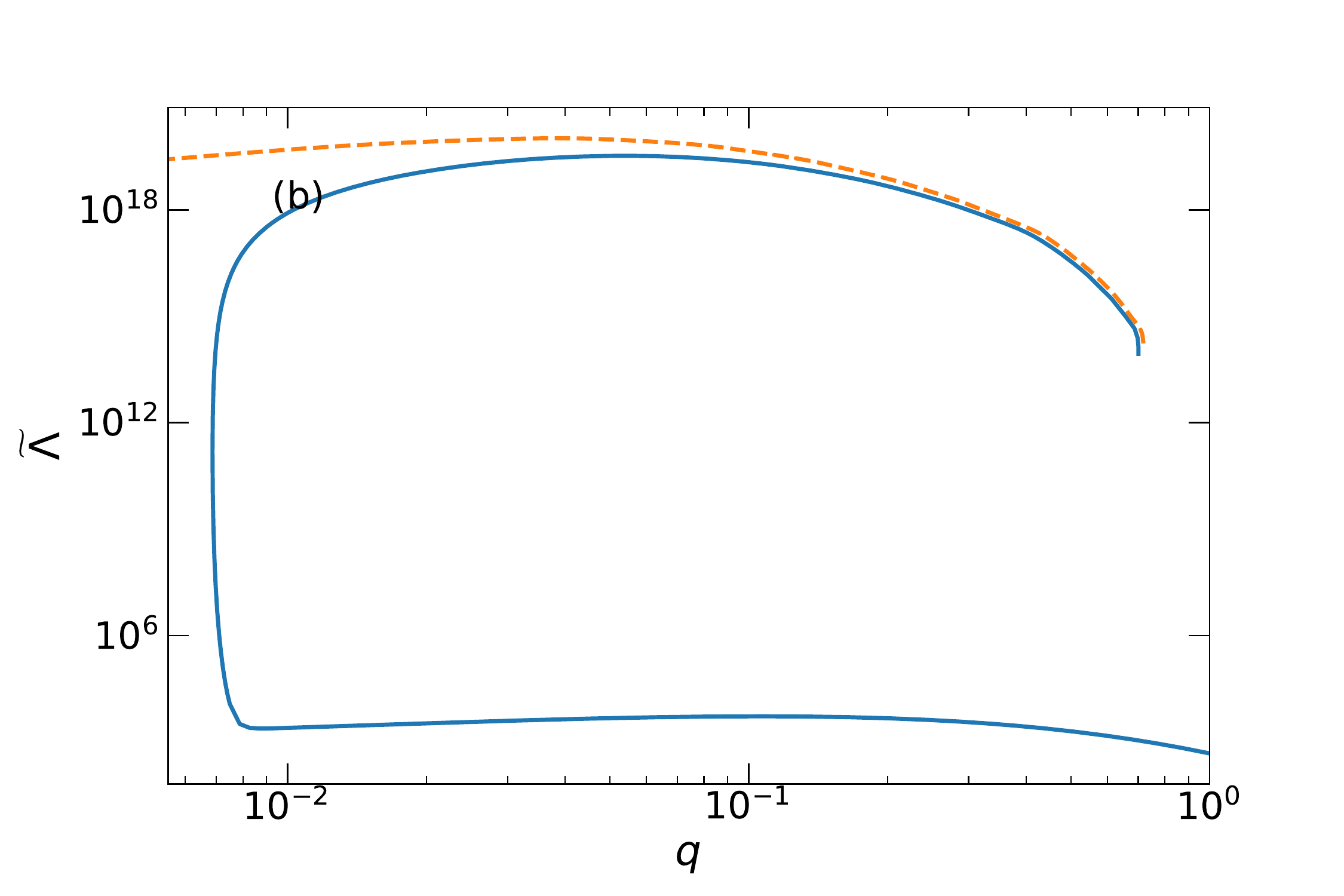}}
\\
\centering
\subfigure{\includegraphics[width=7.5cm]{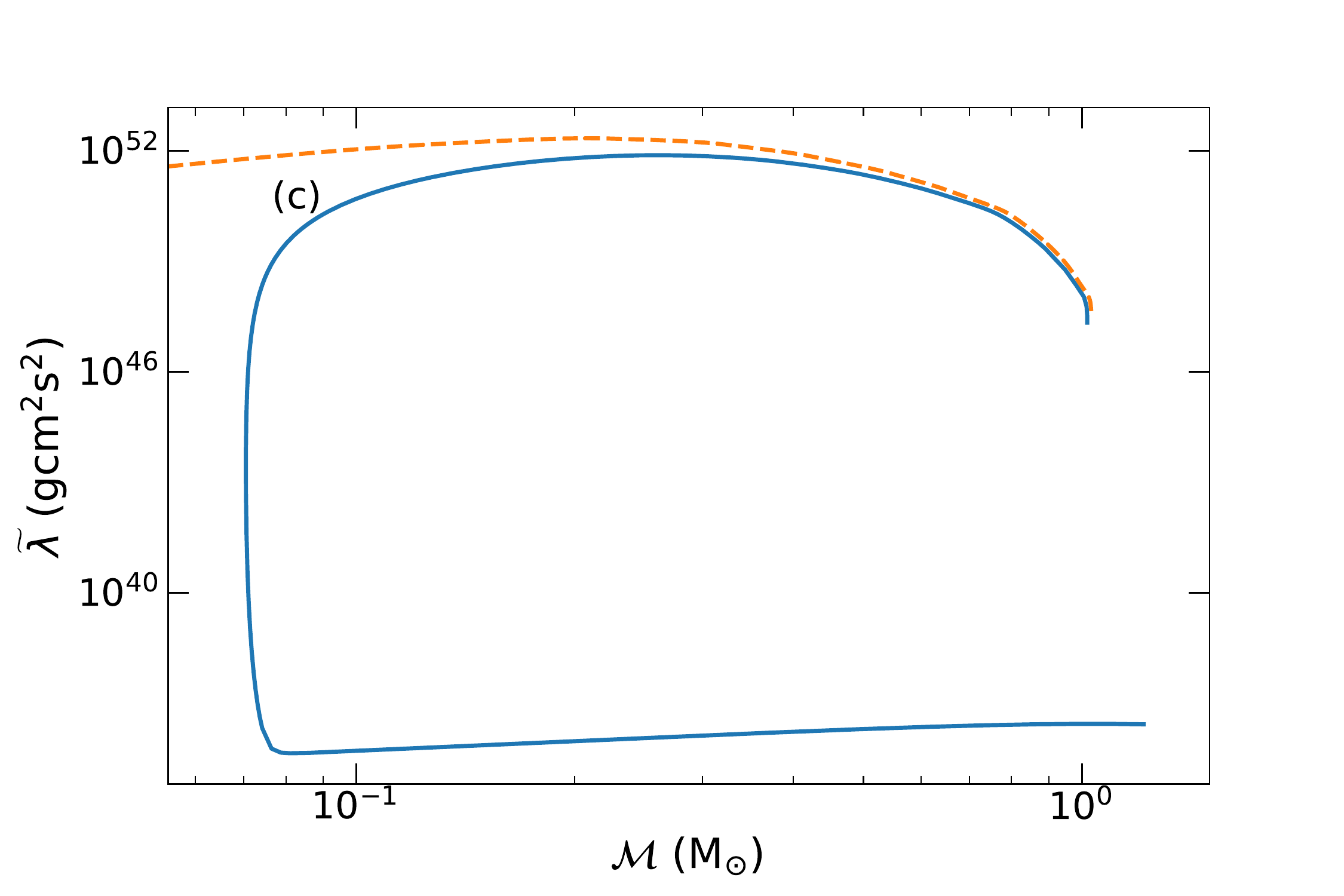}}
\subfigure{\includegraphics[width=7.5cm]{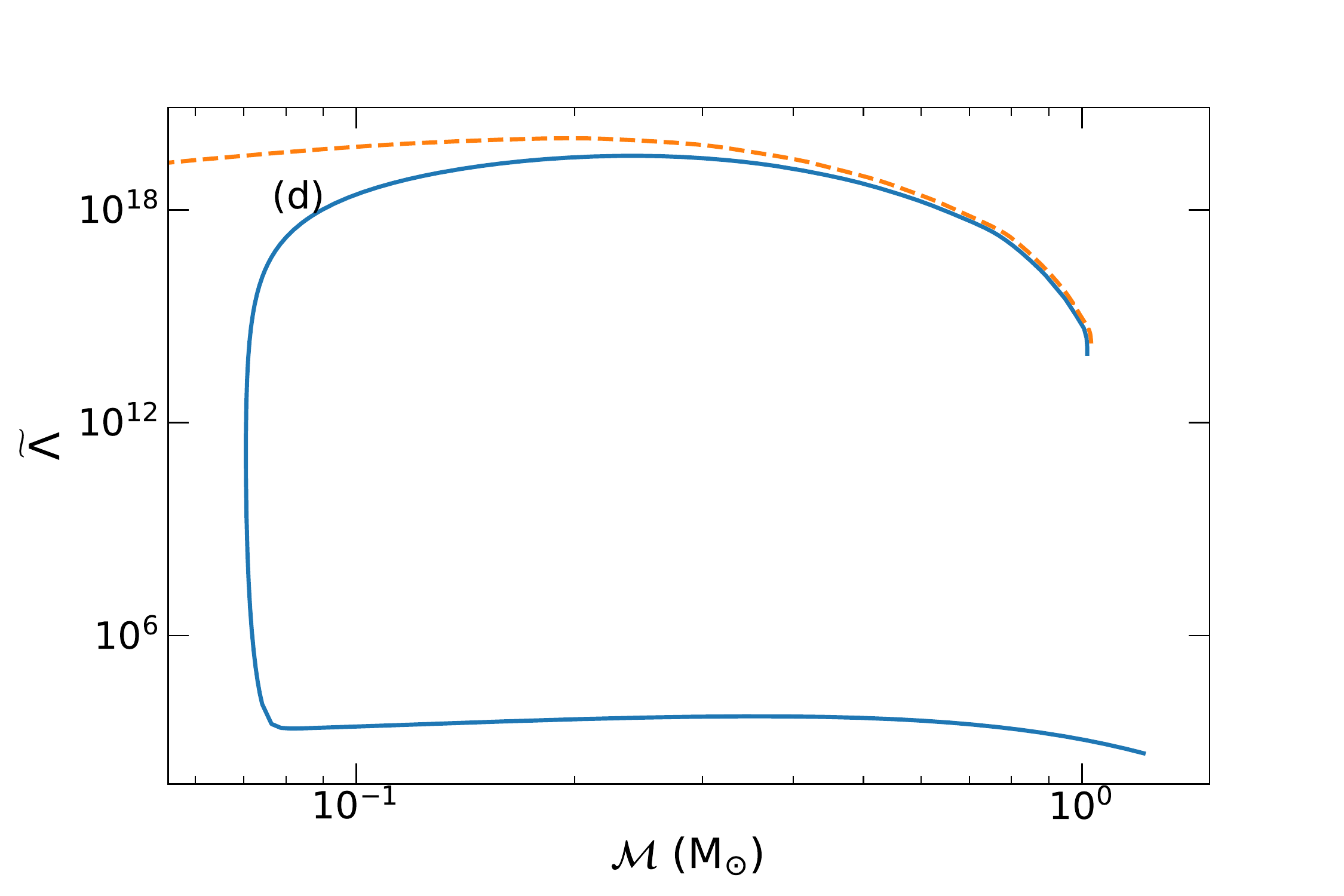}}
\caption{The combined tidal deformability (a) and combined dimensionless tidal deformability (b) as a
   function of the mass radio $q$ for strange quark objects with crust. In Panels (c) and (d), the
   combined tidal deformability and combined dimensionless tidal deformability are plot versus
   the chirp mass $\mathcal{M}$. For comparison, the dashed curve illustrates the case of
   pulsar-white dwarf/normal planet systems. In all the plot, the primary compact star is
   assumed to have a mass of 1.4 M$_{\odot}$.
   For clarity, the unstable branch is not shown in this plot.} 
\label{fig:bc}
\end{figure}

Fig. \ref{fig:bc} illustrates our results on $\widetilde{\lambda}$ and
$\widetilde{\Lambda}$ for strange quark stars with crust. In this figure, both the
primary star (assumed to be of 1.4 M$_{\odot}$) and the smaller companion are covered
by a crust. Since we are mainly interested in the gravitational wave features of
strange star-strange planet systems, we also plot the combined tidal deformability
of compact star-normal white dwarf/normal planet systems for comparison. From this
figure, we can draw at least two conclusions. First, we
see that the combined tidal deformability of the strange star-strange planet/strange
dwarf systems now is generally in the range of  $\rm10^{49}\,g\,cm^2\,s^2$ ---
$\rm10^{51}\,g\,cm^2\,s^2$, it is much larger than that of binary neutron star systems
(which is $ \rm\sim 10^{36}\,g\,cm^2\,s^2$).
This is very different from the case of bare strange quark objects, but it is easy to
understand. When the strange planet/dwarf has a crust, its radius will be significantly
increased so that it will be much sensitive to the tidal effect. Second, when the
companion mass is larger than $\sim 0.1$ M$_{\odot}$, the difference between
a strange star-strange dwarf system and a neutron star-white dwarf system is not significant.
However, when the companion mass is less than $\sim 0.1$ M$_{\odot}$, the difference
between the solid curve and the dashed curve becomes very significant.

\begin{figure}
\centering
\subfigure{\includegraphics[width=7.5cm]{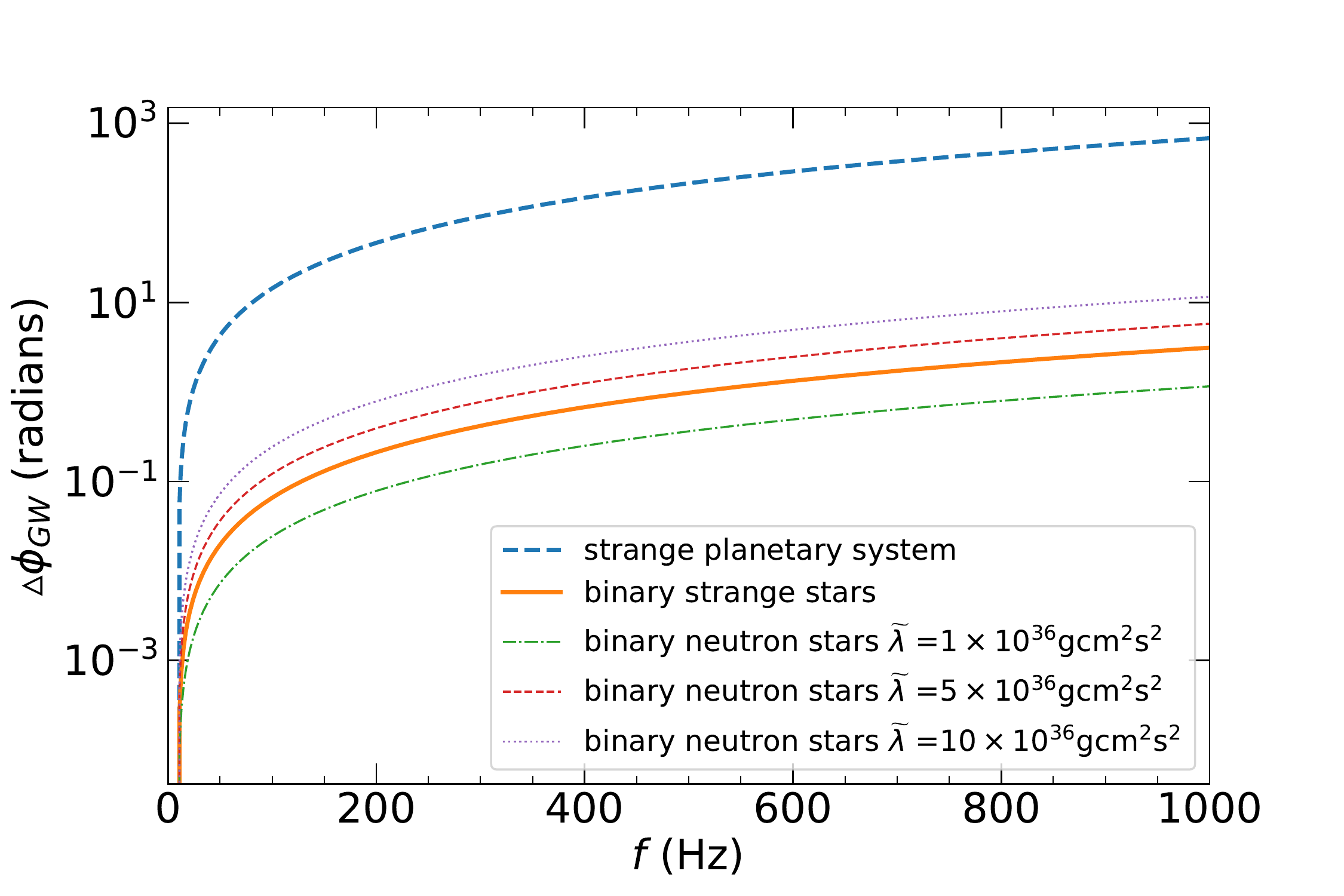}}
\subfigure{\includegraphics[width=7.5cm]{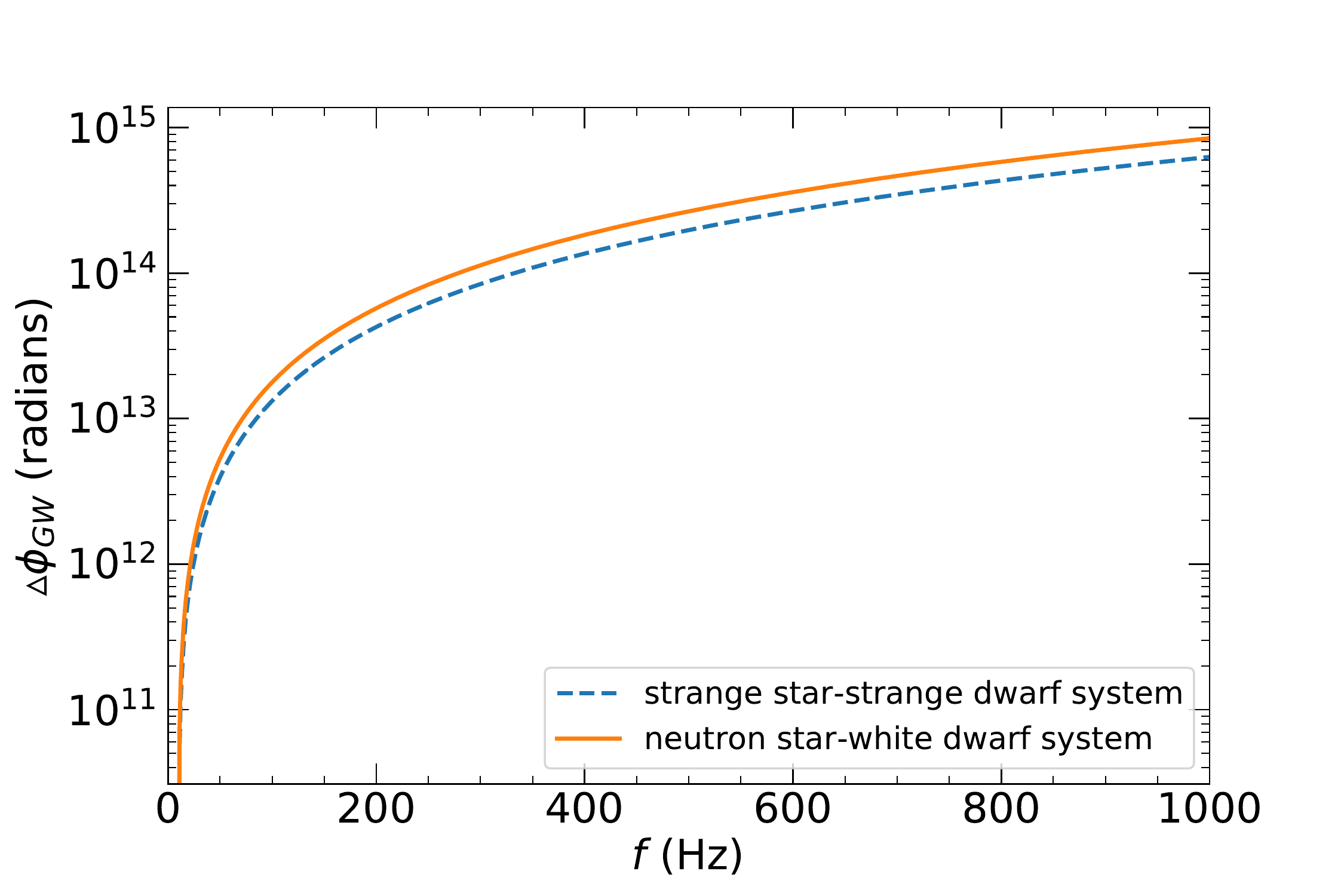}}
\caption{The accumulated phase versus gravitational wave frequency for binary strange stars
   and strange planetary systems (the upper panel). The lower panel shows the case for a strange
   star-strange dwarf binary. For comparison, the accumulated phase of binary neutron star systems
   (with different combined tidal deformability, which corresponds to different EoS) and neutron
   star-white dwarf systems is also plot correspondingly.
   The calculation of the accumulated phase starts at 10 Hz in all the cases.}
\label{fig:pt}
\end{figure}

In order to show the effect of tidal deformation on the gravitational waveform
signal, we calculate the accumulated phase of different systems by \citep{2010PhRvD..81l3016H}
\begin{equation}
    \frac{d\Phi}{dx}=-\frac{195c^{15/3}}{8G^{7/3}}\frac{x^{3/2}\widetilde{\lambda}}{M^5\eta},
\end{equation}
where $x=(\omega M)^{2/3}$ is a dimensionless post-Newtonian parameter, $\omega$ is the orbital angular
frequency and $\eta=m_1m_2/M^2$ is the symmetric mass ratio. The accumulated phases of some representative
systems are considered, including a binary strange star
system ($m_1=m_2=1.4{\rm\,M_{\odot}},\,\widetilde{\lambda}=2.7 \times \rm10^{36}\,g\,cm^2\,s^2$),
a strange star-strange planet system
($m_1=1.4{\rm\,M_{\odot}},\,m_2=1.0{\rm\,M_{jup}},\,\widetilde{\lambda}=1.7 \times \rm10^{35}\,g\,cm^2\,s^2$),
and a strange star-strange dwarf
system ($m_1=1.4{\rm\,M_{\odot}},\,m_2=0.6{\rm\,M_{\odot}},\,\widetilde{\lambda}=1.5 \times \rm10^{50}\,g\,cm^2\,s^2$).
Three different binary neutron star systems
($m_1=m_2=1.4{\rm\,M_{\odot}},\,\widetilde{\lambda}_1=1 \times \rm10^{36}\,g\,cm^2\,s^2,
\widetilde{\lambda}_2=5 \times \rm10^{36}\,g\,cm^2\,s^2,
\widetilde{\lambda}_3=10 \times \rm10^{36}\,g\,cm^2\,s^2$, note that the different $\widetilde{\lambda}_i$ here
corresponds to different EoS of neutron stars) and a neutron star-white dwarf
system ($m_1=1.4{\rm\,M_{\odot}},\,m_2=0.6{\rm\,M_{\odot}},\,\widetilde{\lambda}=2.0 \times \rm10^{50}\,g\,cm^2\,s^2$)
are also calculated for comparison. Our numerical results are shown in Fig. \ref{fig:pt}.

From Fig. \ref{fig:pt}, we see that different EoS generally leads to different accumulated phase.
However, the phase difference between binary neutron stars and binary strange stars (1 -- 2 M$_\odot$) is basically
very small. It would be difficult to discriminate between them by means of gravitational waveform
observations. For strange planetary systems, the accumulated phase is significantly different. Note
that an ordinary planet orbiting around a compact star is not an effective gravitational wave emitter,
because it would be tidally disrupted by the compact host when it gets close enough. Fig. \ref{fig:pt}
further proves the idea that gravitational wave is an effective probe of strange quark
planets \cite{2015ApJ...804...21G,2019AIPC.2127b0027K}. From the lower panel, we see that the
accumulated phase of a neutron star-white dwarf binary is about 1.34 times of that of a strange
star-strange dwarf system. Such a phase difference may hopefully be measured by the next
generation gravitational wave detectors.

\section{Conclusion}\label{sec4}
In this study, we calculate the tidal deformability of strange quark stars, which
could be either bare ones or covered by a normal matter crust. Especially, we
concentrate on less massive objects, i.e., strange dwarfs and strange planets.
For bare strange quark planets, since the internal density is almost constant,
it is found that the tidal deformability simply scales
as $ \lambda \propto m^{5/3}$, while the dimensionless tidal deformability
scales as $\Lambda  \propto m^{-10/3}$. As for the combined tidal deformability,
it is shown that this binary parameter is generally very small for
strange planetary systems, of the order of $\rm 10^{34}\,g\,cm^2\,s^2$, which is much smaller than
that of binary neutron star systems (which is $\rm \sim 10^{36}\,g\,cm^2\,s^2$).
When the strange quark objects have a crust, it is found that the tidal deformability of
strange dwarfs and strange planets are still less than that of their normal matter
counterparts. The difference is especially significant for planetary objects.
It means that strange quark planets will be less deformed in a tidal field as compared
with a normal planet.

Our study clearly shows that it is a hopeful method to try to identify strange quark
object by searching for strange quark planets through gravitational wave observations.
Tidal deformability will be a useful tool to probe the internal structure of compact stars.
In fact, it has been argued that the gravitational wave signals of merging strange star
planetary systems in our Galaxy can be detected by advanced-LIGO.
Encouragingly, in the future, such signals from local galaxies up to several Mpc will even be detectable
to the Einstein Telescope \cite{2015ApJ...804...21G,2019AIPC.2127b0027K}.

It is an interesting problem that how strange quark planets/dwarfs can be formed in the Universe.
Theoretically, there are at least two possibilities. First, they may be born together with their host,
i.e. the primary strange quark star. The primary strange quark star itself may be produced due to an
accretion-induced phase transition of a neutron star, or during a binary neutron star merger.
During these fierce processes, a lot of small chunks of quark matter, which is called ``strangelets'',
will be ejected. They will contaminate surrounding ordinary planets/white dwarfs and produce strange quark
planets/dwarfs. Furthermore, since the newly formed strange quark star is fast spinning and highly turbulent,
it may also eject some large clumps (of planetary mass) to directly produce strange quark planets
around it\citep{2003ChPhL..20..806X,2012RAA....12..813H}. Second, there is a stage in the early Universe
called the quark-gluon phase. At this stage, denser clumps of planetary mass or dwarf mass may form directly.
They may survive to the present and be captured by other stars to form planetary systems
or strange dwarf binaries \citep{1994PhRvL..73.1328C}.

\section*{Acknowledgements}
We thank the anonymous referee for helpful comments and suggestions that lead to an overall improvement of this study.
This work is supported by National SKA Program of China No. 2020SKA0120300, by the National
Natural Science Foundation of China (Grant Nos. 11873030, 12041306, U1938201, 12103055, 11903019, 11833003), 
and by the science research grants from the China Manned Space Project with NO. CMS-CSST-2021-B11.

%PRD style
%ZEncoding:latex%ZLinelength:0\bibitem{%1M%Y} %5I, %J, %V, %p (%Y).\n

\end{document}